\documentclass[aps,prl,reprint,superscriptaddress]{revtex4-1}
\usepackage{amsmath}
\usepackage{graphicx}
\usepackage{braket}

\begin{document}

\title{Possible hundredfold enhancement in the direct magnetic coupling of a single atomic spin to a circuit resonator}


\author{Bahman Sarabi}
\affiliation{National Institute of Standards and Technology, Gaithersburg, MD 20899, USA}
\affiliation{Joint Quantum Institute, University of Maryland, College Park, Maryland 20742, USA}
\author{Peihao Huang}
\affiliation{National Institute of Standards and Technology, Gaithersburg, MD 20899, USA}
\affiliation{Joint Quantum Institute, University of Maryland, College Park, Maryland 20742, USA}
\author{Neil M. Zimmerman}
\affiliation{National Institute of Standards and Technology, Gaithersburg, MD 20899, USA}


\date{\today}

\begin{abstract}
We report on the challenges and limitations of direct coupling of the magnetic field from a circuit resonator to an electron spin bound to a donor potential. We propose a device consisting of a trilayer lumped-element superconducting resonator and a single donor implanted in enriched $^{28}$Si. The resonator impedance is significantly smaller than the practically achievable limit using prevalent coplanar resonators. Furthermore, the resonator includes a nano-scale spiral inductor to spatially focus the magnetic field from the photons at the location of the implanted donor. The design promises approximately two orders of magnitude increase in the local magnetic field, and thus the spin to photon coupling rate $g$, compared to the estimated coupling rate to the magnetic field of coplanar transmission-line resonators. We show that by using niobium (aluminum) as the resonator's superconductor and a single phosphorous (bismuth) atom as the donor, a coupling rate of $g/2\pi$=0.24 MHz (0.39 MHz) can be achieved in the single photon regime. For this hybrid cavity quantum electrodynamic system, such enhancement in $g$ is sufficient to enter the strong coupling regime. 
\end{abstract}


\maketitle

\section{I. Introduction}

Silicon-based spin qubits, including gate-defined quantum dot \cite{RevModPhys.79.1217,PhysRevA.57.120,petta2005coherent} and single-atom \cite{RevModPhys.85.961,pla2012single} devices, use the spin degree of freedom to store and process quantum information, and are promising candidates for future quantum electronic circuits. The electronic or nuclear spin is well decoupled from the noisy environment, resulting in extremely long coherence times \cite{saeedi2013room, tyryshkin2012electron,Steger1280} desirable for fault-tolerant quantum computing. Single-atom spin qubits offer additional advantages over quantum dot qubits such as longer coherence due to strong confinement potentials, and are expected to have intrinsically better reproducibility. Therefore, it is no surprise that silicon-based single-atom spin qubits hold the record coherence times of any solid state single qubit \cite{Muhonen2014}. However, this attractive isolation from sources of decoherence comes at the price of relatively poor coupling to the control and readout units. This causes relatively long qubit initialization times, degraded readout fidelity, and weak spin-spin coupling for multi-qubit gate operations \cite{Awschalom1174}.

One simple way to enhance the coupling rate is to increase the ac magnetic field from the external circuit. In this regard, superconducting circuit resonators are attractive due to their relatively large quality factors, ease of coupling to other circuits, their capability of generating relatively large ac magnetic fields by carrying relatively large currents, and monolithic integration with semiconductor devices.

Over the last two decades, superconducting microwave resonators have had extensive applications that range from superconducting qubit initialization, manipulation and readout \cite{wallraff2004strong, GU20171}, and inter-qubit coupling \cite{majer2007coupling} to dielectric characterization \cite{PhysRevLett.116.167002}. One of the most commonly used superconducting quantum computing architectures is one based on cavity quantum electrodynamics (cQED) \cite{PhysRevA.69.062320}, in which a 2D (circuit based) or 3D cavity is employed to initialize, manipulate and readout the superconducting qubit. Superconducting circuit cavities have not yet found a similar prevalence in spin qubit circuits due to the fact that the magnetic field of a typical superconducting resonator and the spin magnetic-dipole moment are relatively small, leading to an insufficient spin-photon coupling strength for practical purposes. The direct magnetic coupling of coplanar resonators (typically coplanar waveguide resonators) to donor electrons in silicon \cite{PhysRevLett.118.037701,PhysRevLett.102.083602, tosi2014circuit} and diamond nitrogen-vacancy centers \cite{PhysRevLett.105.140502, PhysRevLett.107.060502} can only achieve a maximum single-spin coupling rate of a few kHz.

Several methods have been proposed to enhance the coupling of a single spin to a photon within a superconducting circuit resonator. It is easier to couple a photon to quantum dot spin qubits than to single-atom spins because in the former, the spin dynamics can be translated into an electric dipole interacting with resonator's electric field \cite{petersson2012circuit,Viennot408,PhysRevB.86.035314}. Such architectures, however, require hybridization of the spin states with charge states, coupling charge noise to the spin and thus affecting its coherence.  For instance, in two recent experimental demonstrations of strong spin-photon coupling in Si \cite{mi2018coherent,Samkharadze1123}, the maximum value of the figure of merit ratio of the coupling rate to the spin decoherence rate was about five, due in part to the approximately factor of five increase in the dephasing rate \cite{Samkharadze1123} arising from the hybridization of spin and charge states. 
For the single-atom spin qubits, indirect coupling to a resonator via superconducting qubits \cite{PhysRevLett.107.220501, PhysRevB.81.241202} can enhance the coupling, but imposes nonlinearity on the circuit complicating cQED analysis, and also introduces loss from Josephson junction tunnel barriers \cite{PhysRevLett.93.077003,PhysRevLett.95.210503} and magnetic flux noise \cite{kumar2016origin}.  

\begin{figure*}
\includegraphics[width=\textwidth]{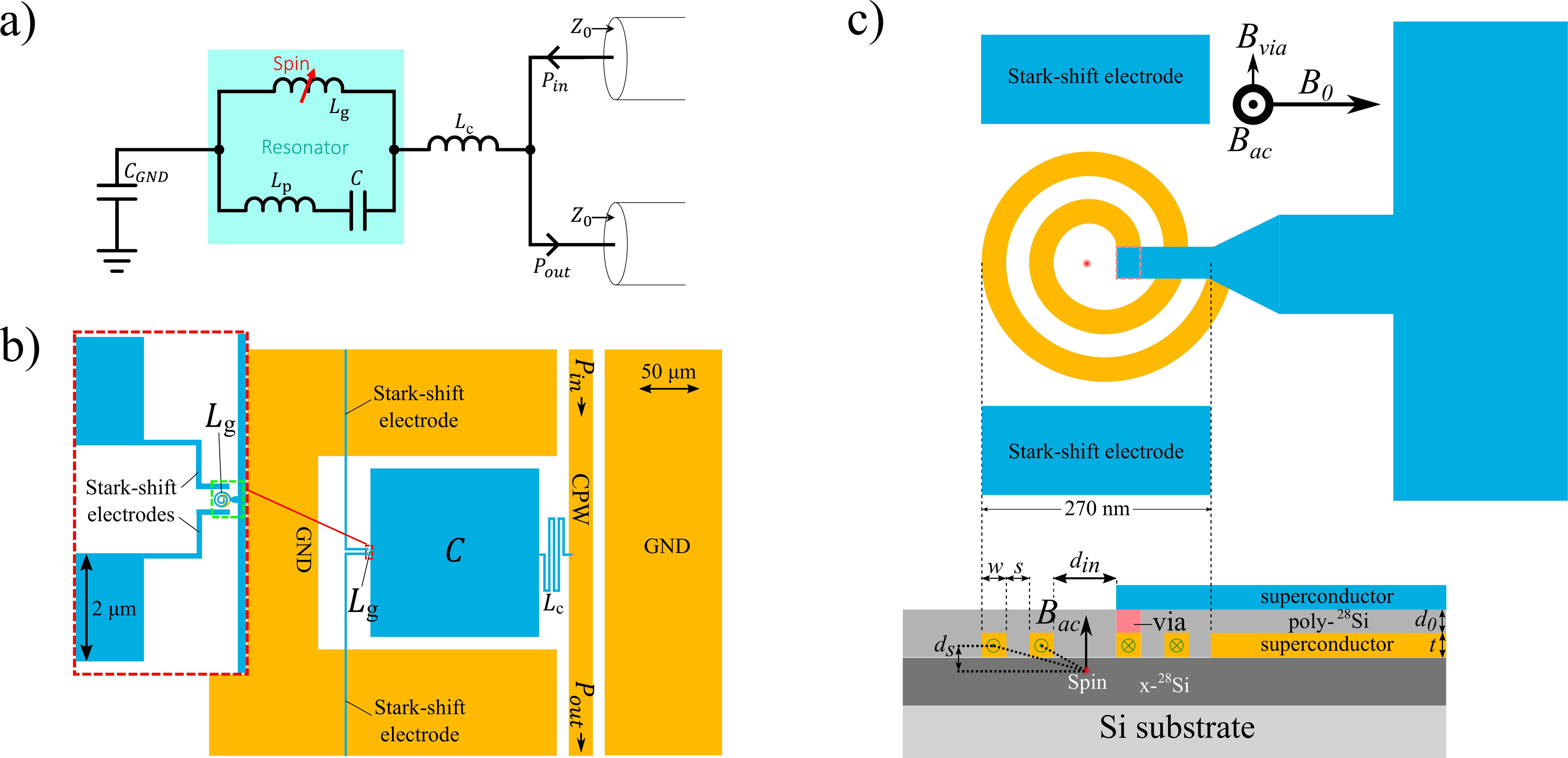}
\caption{(a) Schematic of the circuit showing the resonator galvanically coupled to the coplanar waveguide (CPW). $C_{\mathrm{GND}}$ is the resonator's capacitance to ground. $P_{\mathrm{in}}$, $P_{\mathrm{out}}$, $L_{\mathrm{g}}$ and $L_{\mathrm{p}}$ show the microwave input power, output power, geometric inductance and the parasitic inductance of the resonator circuit, respectively. 
(b) Layout of the device showing resonator's capacitor $C$ and inductor $L$. Blue and orange show the top and bottom superconducting layers, respectively. The area within the red dashed square is magnified. The spiral inductor within the green dashed square is further magnified to show (c) the single-spiral geometry (red dot represents the spin). Dimensions are $w=s=t=d_{in}/2=30$ nm, compatible with standard electron beam lithography, and $d_{0}=d_{\mathrm{s}}=25$ nm. The crystalline $^{28}$Si (x-$^{28}$Si) growth is required only in the volume surrounding the spin, and amorphous or polycrystalline Si is acceptable everywhere else. The relatively weak magnetic field $\mathbf{B}_{\mathrm{via}}$ (directions shown represent magnetic fields at the location of the spin) from the via determines the best choice for the direction of the static magnetic field, $\mathbf{B}_{0}\parallel(\mathbf{B}_{\mathrm{via}}\times\mathbf{B}_{\mathrm{ac}})$, where $\mathbf{B}_{\mathrm{ac}}$ is the magnetic field from the spiral inductor. This ensures that the total ac field is perpendicular to $\mathbf{B}_{0}$. The current flowing through the spiral inductor is shown by green in- or out of plane vectors.}
\label{layout}
\end{figure*}

In this paper, we show that replacing the coplanar transmission line resonator with a lumped-element circuit resonator that includes a spiral inductor, can lead to a dramatic enhancement in the spin-photon magnetic coupling rate $g$ of approximately two orders of magnitude. 
As we will discuss, this improvement is a result of the reduced resonator impedance thanks to the trilayer lumped-element design, as well as employing nano-scale spiral inductor loops which effectively localize the resonator's magnetic field at the location of the spin, eliminating the need for an Oersted line \cite{Lauchte1500022} or a micrometer-scale magnet (micromagnet) \cite{PhysRevLett.96.047202,mi2018coherent}. We also show that this coupling rate is enough to take the spin-resonator hybrid system to the strong coupling regime, where $g$ is larger than or approximately equal to the resonator decay rate $\kappa$. 
In order to achieve this relatively large $g$, we need a relatively small total inductance $L_\mathrm{tot}$, and concomitantly small impedance $Z=\sqrt{L_\mathrm{tot}/C}$. To reach the desired operation frequency $\omega_0=\sqrt{1/L_\mathrm{tot}C}$, we need a large capacitance $C$, and thus coplanar interdigitated capacitances (IDCs) are practically insufficient. Furthermore, IDCs generally introduce relatively large parasitic inductance which could set a limit to the minimum achievable $Z$. Therefore, the proposed device geometry, compatible with standard micro- and nanofabrication techniques, includes a trilayer (parallel-plate) capacitor with a deposited insulating layer. Trilayer capacitors, however, give rise to a lower resonator quality factor $Q$ than that of a typical coplanar geometry. Nevertheless, as our calculations show, the increase in $g$ resulted from employing a trilayer lumped-element design is large enough to overcome the limited $Q$, allowing for strong spin-photon coupling.
\section{II. Single-atom device}
The lossless dynamics of a spin-$\frac{1}{2}$ system with spin transition frequency $\omega_{\mathrm{s}}$ coupled with rate $g$ to a cavity resonant at $\omega_{0}$, are described by the Jaynes-Cummings Hamiltonian, $\mathcal{H}_\mathrm{JC}=\hbar\omega_{0}(\hat{a}^{\dagger}\hat{a}+1/2)+(1/2)\hbar\omega_{\mathrm{s}}\hat{\sigma}^{z}+\hbar g(\hat{a}^{\dagger}\hat{\sigma}^{-}+\hat{\sigma}^{+}\hat{a})$ \cite{1443594}. Here, $\hat{a}$ and $\hat{\sigma}$ are photon and spin operators, respectively. The spin-photon magnetic coupling rate is obtained as $g=g_{\mathrm{e}}\mu_{\mathrm{B}}B_{\mathrm{ac},0}\braket{\mathrm{g}|\hat{S}_{x}|\mathrm{e}}/\hbar$ where $g_{\mathrm{e}}\simeq2$ is the electron g-factor, $\mu_{\mathrm{B}}$ is the Bohr magneton, and $B_{\mathrm{ac},0}=\braket{n=0|\hat{B}_{\mathrm{ac}}^{2}|n=0}^{1/2}$ is the root mean square (RMS) local magnetic field at the location of the spin with $n=0$ photons on resonance. $\ket{\mathrm{g}}$ and $\ket{\mathrm{e}}$ are the ground and excited spin states, respectively, that couple to the microwave field; for spin-$\frac{1}{2}$, $\braket{\mathrm{g}|\hat{S}_{x}|\mathrm{e}}=1/2$. Finally, the spin rotation speed can be enhanced with a larger local RMS magnetic field $B_{\mathrm{ac}}=\braket{n|\hat{B}_{\mathrm{ac}}^{2}|n}^{1/2}$ for any arbitrary $n$.

The schematic and layout of the proposed circuit are shown in Figs. \ref{layout}(a) and \ref{layout}(b-c), respectively, where a LC resonator is coupled to a coplanar waveguide (CPW) using a direct (galvanic) connection through a coupling inductor $L_\mathrm{c}$ and the donor is within the resonator's deliberate (geometric) inductor $L_\mathrm{g}$. The galvanic coupling, employed in previous experiments \cite{vissers2015frequency}, can help to achieve the desired CPW-resonator coupling rates especially when the resonator impedance is significantly different from the CPW's characteristic impedance $Z_{0}=50\ \Omega$.   
The capacitance $C$ is provided using a trilayer capacitor. The RMS current through the inductor at an average photon number $\bar{n}$ on resonance is $I_{\mathrm{ac}}=\sqrt{(\bar{n}+1/2)\hbar\omega_{0}/L_{\mathrm{tot}}}$, where $\omega_{0}$ is the resonance frequency and $L_{\mathrm{tot}}$ denotes the total inductance within the resonator circuit. Since the desired ac magnetic field from the spiral loop(s) is proportional to $I_{\mathrm{ac}}$, it is clear that the inductance (or impedance $Z$) must be minimized and the photon frequency must be maximized to obtain the maximum magnetic field. Therefore, it is necessary to study the sources of inductance and obtain operation frequency limitations. 

\begin{figure}
	\includegraphics[width=3 in]{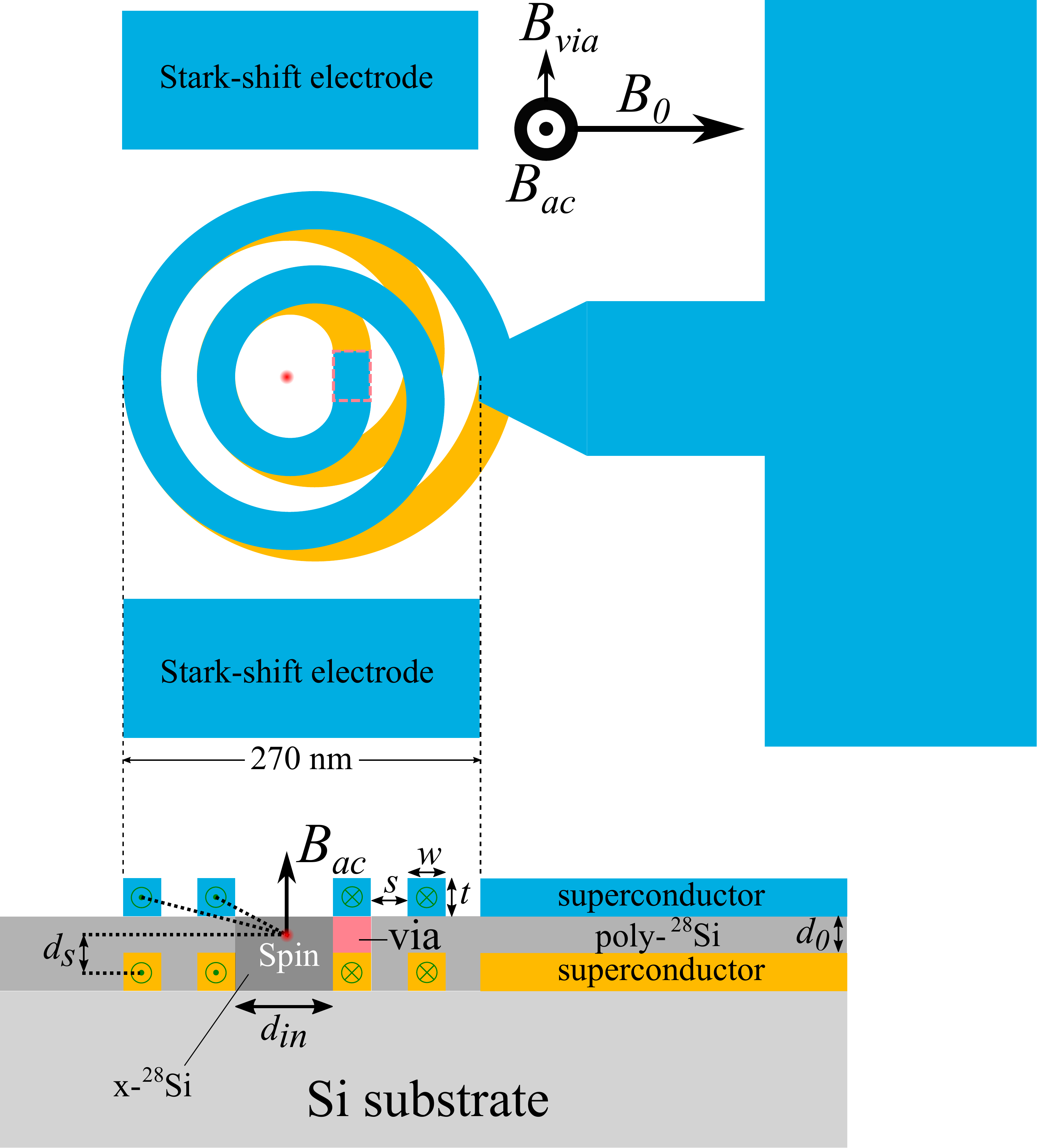}
	\caption{Layout of the double-spiral (2S) device in the vicinity of the nanoscale spiral inductor. Here, $d_{\mathrm{s}}=27.5$ nm and other dimensions are identical to those in Fig. \ref{layout}(c).}
	\label{2slayout}
\end{figure}

We study two sets of materials, one that uses niobium as superconductor and phosphorous as donor (Nb/P), and one with aluminum as superconductor and bismuth as donor (Al/Bi). Fabrication is considered to be slightly better established with Al, whereas Nb has a much higher critical magnetic field of $B_{\mathrm{c}1,\mathrm{Nb}}=0.2$ T, which is the highest $B_{\mathrm{c}1}$ among the single-element superconductors. Through the Zeeman effect, for the Nb/P set, the uncoupled electron spin-up ($\ket{\uparrow}$) and spin-down ($\ket{\downarrow}$) states are split by the photon frequency $\omega_0/2\pi=5.6$ GHz at the maximum $B_{0}=\hbar\omega_{0}/g_{\mathrm{e}}\mu_{\mathrm{B}}\simeq B_{\mathrm{c}1,\mathrm{Nb}}$, not to quench superconductivity. For the Al/Bi set, we consider operating at $B_{0}<10$ mT (below the critical magnetic field of Al), and use the splitting of the spin multiplets ($\omega_{0}/2\pi=7.375$ GHz) of the Bi donor. These splittings arise from the strong hyperfine interaction between the electron ($S=1/2$) and Bi nuclear spin ($I=9/2$) \cite{PhysRevB.86.245301}, leading to a total spin $F$ and its projection $m_{F}$ along $\mathbf{B_{0}}$. By employing the $\ket{F,m_{F}}=\ket{5,-5}\leftrightarrow\ket{4,-4}$ transition corresponding to the largest $\hat{S}_{x}$ matrix element (0.47), and a static magnetic field of $B_{0}=5$ mT, the multiplet degeneracy at $B_{0}=0$ is lifted by more than 20 MHz \cite{Bienfait2016,PhysRevLett.105.067602}, enough to decouple the nearby transitions.

\begin{figure}
	\includegraphics[width=\linewidth]{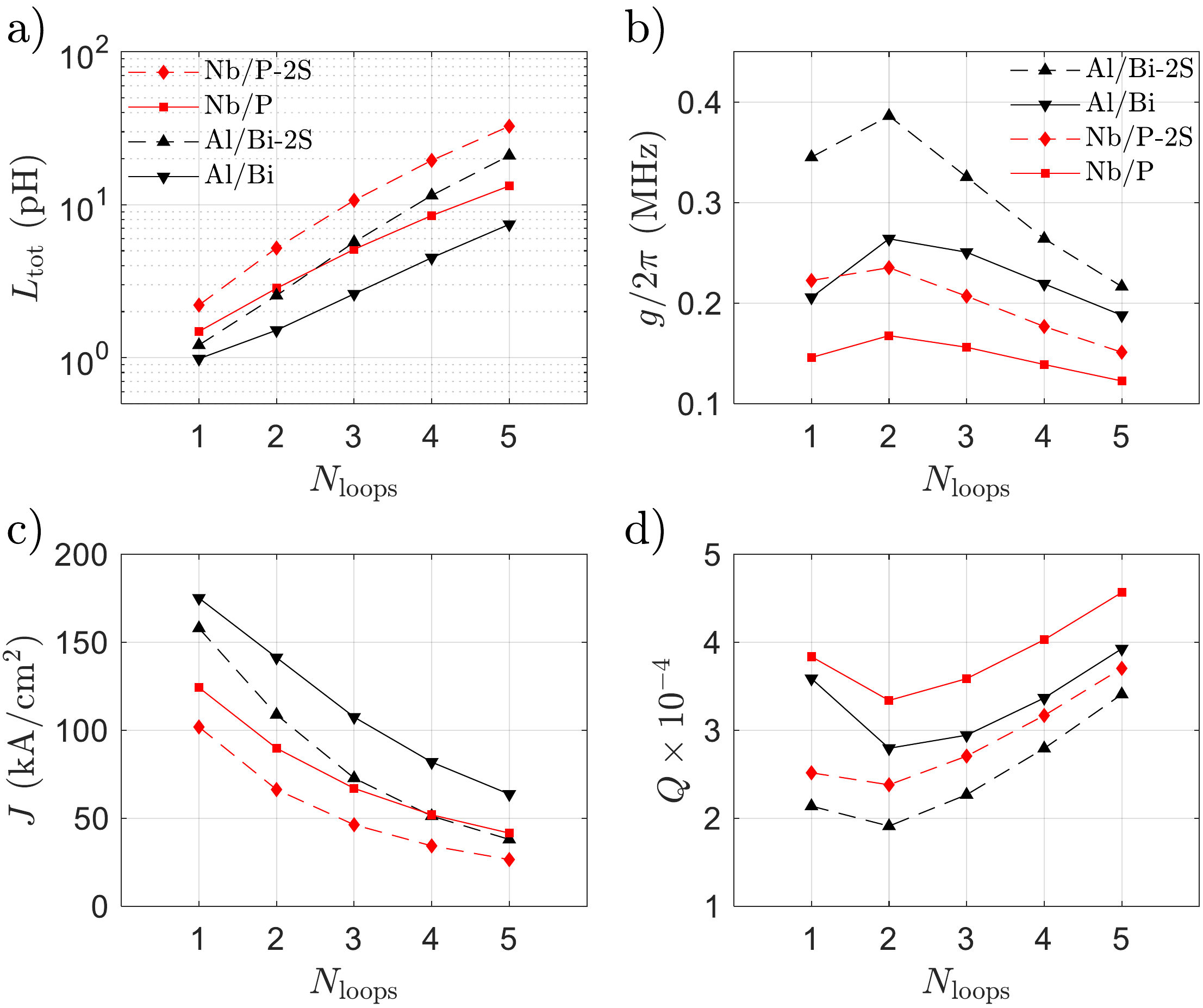}
	\caption{Plots of the (a) total inductance, 
		(b) spin-photon coupling rate, (c) vacuum fluctuations current density, and 
		(d) required resonator quality factor $Q\equiv\omega_0/g$ to allow strong spin-photon coupling  
		versus the number of spiral loops $N_{\mathrm{loops}}$ for different combinations of materials and geometries, {\it i.e.} superconducting aluminum and Bismuth donor (Al/Bi), superconducting niobium and phosphorous donor (Nb/P), single-spiral and double-spiral (2S) geometries.}
	\label{plots}
\end{figure}

Figure \ref{2slayout} shows an alternative double-spiral (2S) geometry where the spiral inductor extends to both superconducting layers. Here we see the following optimum combination of materials: {\it i}) Amorphous or polycrystalline Si can be deposited on the metal layers; since they have similar permittivities to crystalline Si, the capacitor size will be similar. {\it ii}) Since the spin does not have a metal layer below it, one can still deposit crystalline $^{28}$Si and then implant the donor, thus avoiding the fast decoherence which would otherwise result from the non-crystalline, non-enriched films \cite{Fuechsle2012}. 
From the fabrication point of view, it is easiest to use the same $^{28}$Si film as the capacitor dielectric (see Figs. \ref{layout}(c) and \ref{2slayout}). 
It is noteworthy that using a trilayer design makes the resonator quality factor effectively independent from the surrounding material as almost all the electric field energy is confined within the capacitor dielectric. This is useful for the integration of single electron devices that require lossy oxide layers.

For all four aforementioned combinations of resonator material and geometry, $L_{\mathrm{tot}}$ is simulated and shown in Fig. \ref{plots}(a) as a function of number of spiral loops, $N_{\mathrm{loops}}$ (see Appendix for the details of inductance calculations and simulations). $N_{\mathrm{loops}}$ refers to the number of loops in a single spiral layer regardless of the geometry, {\it e.g.}, $N_{\mathrm{loops}}=2$ for both Figs. \ref{layout}(c) and \ref{2slayout}. Increasing $N_{\mathrm{loops}}$ from 1 to 2 increases $B_{\mathrm{ac},0}$, but, for $N_{\mathrm{loops}}>2$, the competing effect of larger $L_{\mathrm{tot}}$ due to larger loop radii suppresses $I_{\mathrm{ac}}$ and lowers $g$. This effect is clearly demonstrated in Fig. \ref{plots}(b) as the optimum $N_{\mathrm{loops}}=2$ for both material sets, where the vacuum fluctuation's coupling rates for the Al/Bi and Nb/P configurations are obtained as $g/2\pi=0.26$ MHz and 0.17 MHz, respectively. For the 2S geometry, coupling rates for the Al/Bi and Nb/P configurations approach $g/2\pi=0.39$ MHz and 0.24 MHz, respectively. These values are approximately two orders of magnitude larger than previously proposed architectures that use coplanar transmission line resonators \cite{tosi2014circuit}. This new regime of $g$ can enable new experiments such as individual spin spectroscopy, and as we report in section IV, it can lead to significant improvements in spin qubit performance in regard with qubit initialization, manipulation and readout.

Figure \ref{plots}(c) shows the vacuum fluctuation's current density $J=(1/wt)I_{\mathrm{ac}}|_{\bar{n}=0}$ for each configuration, where $w$ and $t$ are the width and thickness of the spiral traces, respectively (see Figs. \ref{layout}(c) and \ref{2slayout}). Clearly, $J$ stays far below the critical current values of Al ($J_{\mathrm{c}}\sim10^{4}$ kA/cm$^{2}$ \cite{PhysRevB.26.3648}) and Nb ($J_{\mathrm{c}}\sim10^{3}$ kA/cm$^{2}$ \cite{1058718}).

The condition for the system to enter the strong coupling regime is $Q\gtrsim\omega_{0}/g$ in the single photon regime, where $Q=\omega_{0}/\kappa$ and $\kappa$ are resonator's total quality factor and total photon decay rate, respectively. Thanks to the relatively large spin-photon coupling rate, resonator quality factors in the range of $10^{4}$ are enough to take the system to the strong coupling regime for all four configurations (see Fig. \ref{plots}(d)). In general, in the limit of low drive power and low temperature, trilayer resonators are significantly lossier than coplanar resonators due to the fact that almost all the photon electric energy is stored within the parallel-plate capacitor dielectric which contains atomic-scale defects. These defects act as lossy two-level fluctuators at small powers and low temperatures, and limit the resonator $Q$ \cite{PhysRevLett.95.210503}. However, loss tangents in the vicinity of $\tan\delta_0\simeq10^{-5}$ have been measured for deposited amorphous hydrogenated silicon (a-Si:H) at single photon energies \cite{o2008microwave} promising resonator $Q$'s approaching $10^{5}$. More recently, elastic measurements have indicated the absence of tunneling states in a hydrogen-free amorphous silicon film suggesting the possibility of depositing ``perfect'' silicon \cite{PhysRevLett.113.025503}, promising even higher $Q$ trilayer resonators using silicon as capacitor dielectric.

It has been previously shown that the donor can be ionized in the vicinity of metallic or other conductive structures due to energy band bendings \cite{Fuechsle2012}. For Al-Si interface, the relatively small work function difference of -30 meV (4.08 eV for Al and 4.05 eV for Si) is smaller than the donor electron binding energy of -46 meV \cite{PhysRevB.23.2082}, and hence using aluminum is expected to allow the donor bound state. However, the Nb work function of 4.3 eV causes significant band bending which can result in donor ionization. By biasing the microwave transmission line and hence the resonator through the galvanic connection with a DC voltage, the potential energy landscape in the neighborhood of the donor can be modulated to recreate a bound state. If a conduction band electron is required to fill the bound state, solutions such as shining a light pulse using a light emitting diode to create excess electron-hole pairs or using an ohmic path to controllably inject electrons can be employed.

From Fig. \ref{plots}(d) and the data available from silicon-based low-loss deposited dielectrics, we infer that the proposed single-atom device can achieve the strong coupling regime. In general, the single-spiral design is expected to be easier to fabricate at the expense of smaller $g$ compared to the double-spiral (2S) geometry.

\section{III. Simplified Device with Spin Ensemble}
We now discuss a greatly simplified proof-of-principle device and experiment. We consider a low-impedance trilayer resonator on top of a Bi-doped substrate or $^{28}$Si layer, and propose to measure the spectrum in the single photon regime while the electron spin transitions are tuned across the resonator bandwidth. Similar to the single-atom device, tuning of the spins can be performed using magnetic (Zeeman) or electric (Stark, see Fig. \ref{layout}) fields, or a combination of both. For further simplification, a single inductor loop can be employed with dimensions compatible with standard photolithography. In this simplified experiment with {\it no single-donor implantation or e-beam lithography requirements}, one can choose Al as superconductor and thus the kinetic inductance within the circuit is negligible with micrometer-scale dimensions of the inductor loop. When $N_{s}$ spins are incorporated and exposed to the relatively uniform magnetic field inside the inductor loop, the collective coupling rate from the spin ensemble becomes $g_{\mathrm{col}}=g\sqrt{N_{s}}$ \cite{PhysRevLett.102.083602}. This enhanced coupling rate would yield clearer spin-resonator interaction features in the spectra for this first proof-of-principle version of the device compared to the single-atom device. 

\begin{figure}
	\includegraphics[width=\linewidth]{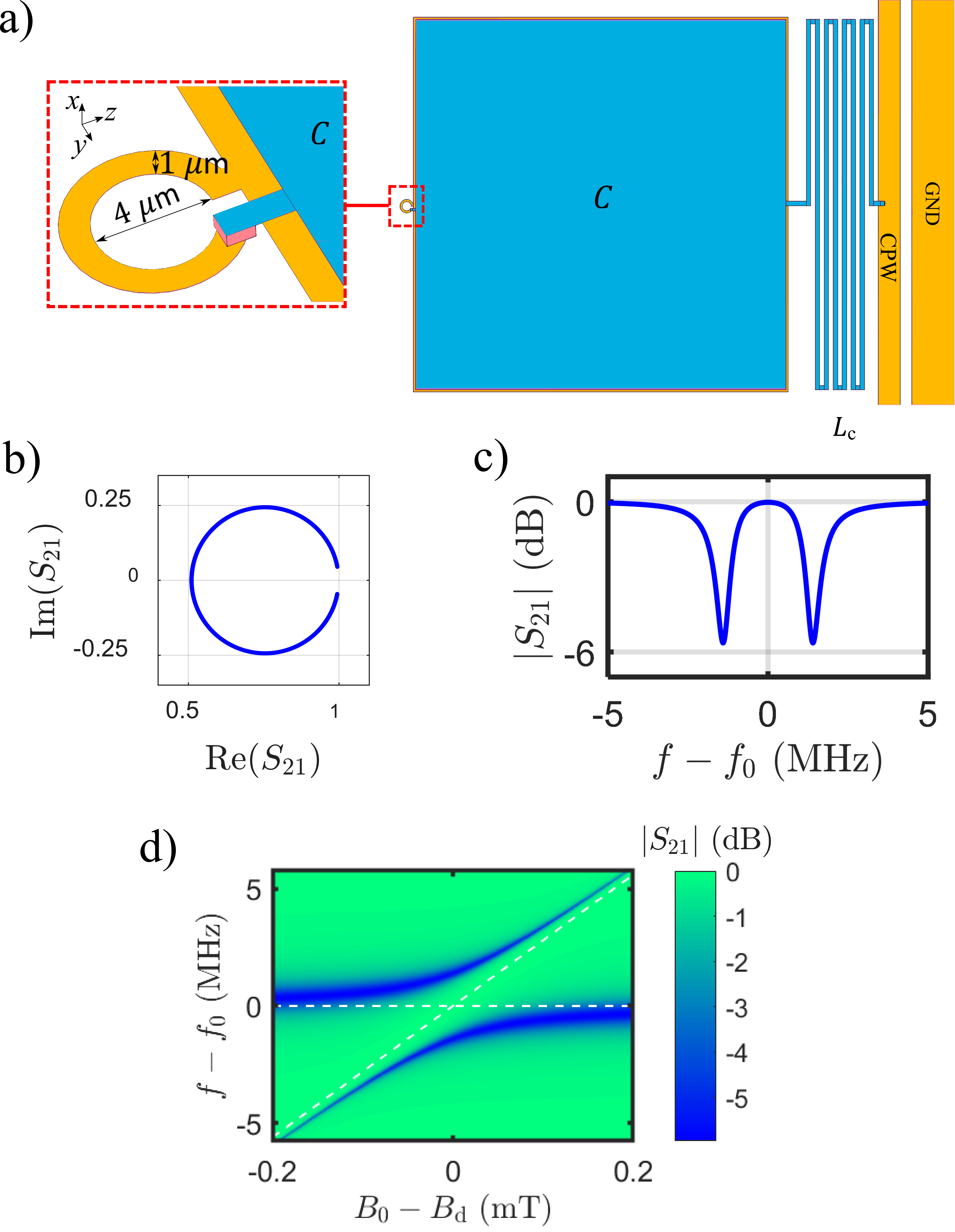}
	\caption{(a) Layout of the simulated Al resonator coupled to an ensemble of Bi spins (ground plane surrounding the resonator and the Stark-shift electrodes are not shown). Here, the dielectric thickness for the capacitor $C$ is assumed to be $d=50$ nm.  (b) Simulated resonator transmission around its resonance frequency $f_0=7.2$ GHz in the absence of spins (no doping). Spectroscopy simulation of the hybrid resonator-spin ensemble device near $f_0$ for (c) zero spin-resonator detuning and (d) with tunable static magnetic field $B_0$ near the degeneracy field $B_\mathrm{d}$.}
	\label{ensemble}
\end{figure}

Since the spin ensemble device is micrometer-scale, high-frequency simulations using electromagnetic simulation software become feasible (Sonnet\textsuperscript{\textregistered} was used for this work). The layout of the simulated device is shown in Fig. \ref{ensemble}(a). We employ a galvanic connection to couple this device to the CPW, similar to the single-atom device described in section II. As shown in Fig. \ref{ensemble}(b), in the absence of spins, the simulated transmission $S_{21}=V_{\mathrm{out}}/V_{\mathrm{in}}$ through the CPW shows a characteristic resonance circle with a diameter of 0.5. This indicates that, because of the galvanic coupling and despite $Z\simeq0.4\ \Omega\ll Z_0$, critical coupling can be achieved, where the resonator's external quality factor $Q_{\mathrm{e}}\equiv\omega_{0}/\kappa_\mathrm{e}$ equals its internal quality factor $Q_{\mathrm{i}}\equiv\omega_{0}/\kappa_\mathrm{i}$. Here $\kappa_\mathrm{e}$ is the resonator's photon loss rate through coupling to the CPW, and $\kappa_\mathrm{i}$ is the resonator decay rate due to it's internal losses, and we have $\kappa=\kappa_\mathrm{e}+\kappa_\mathrm{i}$. In this simulation, we have assumed $\tan\delta_0=10^{-4}$ for the capacitor dielectric.

We adjust $Q_\mathrm{e}$ through simulation; we target $Q_\mathrm{e}=Q_\mathrm{i}$ in order to optimize the coupling while maintaining resonator's coherence by avoiding a large $\kappa_\mathrm{e}$. This critical CPW-resonator coupling regime can not be practically achieved using typical capacitive coupling geometries, and motivated the galvanic coupling scheme. We estimate that in this device, the galvanic connection enhances $\kappa_\mathrm{e}$ by approximately one order of magnitude compared to capacitive coupling with dimensions compatible with standard photolithography.

To obtain $g_{\mathrm{col}}$, we assume that the current within the loop is split equally and flows on the inner and outer edges of the superconducting loop. This is a more realistic flow compared to a single current loop or a uniform current density through the inductor cross section. We consider a cylindrical volume $V$ with uniform doping density of $10^{17}$ cm$^{-3}$ inside the inductor loop with a depth of 500 nm from the substrate surface. The $x$-component of $B_\mathrm{ac}$ from both loops is then calculated at each individual spin location, giving a $g_\mathrm{i}$ for each spin $i$ within $V$. Using an approach described in \cite{bienfait2016reaching}, we find that a histogram of $g_i$ shows a relatively sharp peak at $g_\mathrm{peak}/2\pi=2.2$ kHz and the number of spins within the full width at half maximum is obtained as $4\times10^5$, yielding $g_{\mathrm{col}}/2\pi=1.4$ MHz. The value of $g_\mathrm{peak}$ is significantly smaller than the range of $g$ shown in Fig. \ref{plots}(b) for the single-atom device, because the inductor loop is much larger in the spin ensemble device. 

In order to simulate the device transmission spectroscopy when coupled to the spin ensemble, we consider the calculated value of $g_{\mathrm{col}}$ and use a theoretical model previously developed for a different quantum device with similar physics \cite{sarabi2014cavity}. The model is based on the Jaynes-Cummings model \cite{1443594} and the so-called input-output theory \cite{PhysRevA.30.1386}, with the assumption that the coupled spin ensemble can be treated as an oscillator in the limit of small power and low temperature \cite{:/content/aip/journal/apl/106/17/10.1063/1.4918775}. The theory also treats the CPW as an extremely low-$Q$ ($Q_\mathrm{b}\lesssim1$) cavity with resonance frequency $\omega_\mathrm{b}$ and photon annihilation operator $\hat{b}$. This allows accurate simulation of devices with asymmetric transmission, and is not limited to perfectly symmetric transmissions \cite{sarabi2014cavity}. The Hamiltonian now becomes $\mathcal{H}=\mathcal{H}_\mathrm{JC}+\hbar\omega_\mathrm{b}\hat{b}^\dagger\hat{b}+\hbar c(\hat{b}^\dagger\hat{a}+\hat{a}^\dagger\hat{b})$ where $c$ is the coupling rate between the resonator and cavity $b$, and the spin-related parameters in $\mathcal{H}_\mathrm{JC}$ are now associated with the spin {\it ensemble}. Figures \ref{ensemble}(c-d) show the spectroscopy simulation results using $\tan\delta_0=10^{-4}$ and $Q_\mathrm{i}=Q_\mathrm{e}$ for the resonator with a symmetric transmission, and a conservative value of $T_1=100\ \mu$s for the intrinsic (not Purcell-limited) relaxation time of the spin ensemble. In the presence of experimental spectroscopy data (not presented here), one can extract important parameters of this hybrid spin-resonator system directly using this theory. These parameters include $g_{\mathrm{col}}$, the resonator $Q$ and the intrinsic relaxation time $T_{1}$ for the spin ensemble.


From the high-frequency simulation, the geometric inductance of the loop is extracted as $L_{\mathrm{tot}}=8.5$ pH. 
Despite the significantly larger inductor dimensions of the spin ensemble device, its impedance is similar to that of the single-atom device with a nanoscale spiral inductor for $N_{\mathrm{loops}}=2-3$, primarily due to the smaller contribution of the kinetic inductance in the spin ensemble device. Therefore, we conclude that the CPW can also be sufficiently coupled to the single-atom device using the same (galvanic) coupling scheme.

\section{IV. Qubit operation}

The qubit operation parameters of the single-atom device presented in this paper are adopted from a commonly used cQED approach \cite{PhysRevA.69.062320}. To estimate the initialization, manipulation and readout performance, we focus on the Nb/P-2S configuration with $g/2\pi=0.24$ MHz, $Q_i=4\times10^4$ and $Q_e=4\times10^5$ realizable according to our estimation of capacitive coupling \cite{martinis2014calculation}, or galvanic coupling. We also assume that the bare cavity resonance frequency is $\omega_{0}/2\pi=5.6$ GHz and the Zeeman splitting frequency $\omega_\mathrm{s}$ is tunable around $\omega_{0}$. 

{\bf Initialization:} The zero-detuning ($\Delta\equiv\omega_{\mathrm{s}}-\omega_0=0$) relaxation time limited by the Purcell effect for this strongly-coupled system is obtained as $T_{1,init}=\Gamma_{\mathrm{P}}^{-1}=2\kappa^{-1}=2.3\ \mu$s \cite{PhysRevB.89.104516} which is several orders of magnitude smaller than the free spin relaxation time. 
 The linear dependence of $\Gamma_{\mathrm{P}}$ on $\kappa$ is a result of strong coupling, which distinguishes it from previously measured weakly coupled systems \cite{Bienfait2016} where $\Gamma_{\mathrm{P}}\propto g^2/\kappa$. Also, the 2-order of magnitude increase in $g$ in our device contributes to this dramatically fast initialization.

{\bf Manipulation:} We can Stark shift the spin states using the electrodes shown in Fig. \ref{2slayout}, or use magnetic field to tune the Zeeman energy. For a demonstration of the qubit operation parameters of the device, we choose to operate at $\Delta=40g$, where the strong-coupling Purcell rate becomes $\Gamma_{P}/2\pi=\frac{1}{4\pi}(\kappa-\sqrt{2}\sqrt{-A+\sqrt{A^2+\kappa^2\Delta^2}})=88$ Hz with $A\equiv\Delta^2+4g^2-\kappa^2/4$ \cite{PhysRevB.89.104516}, giving rise to $\mathrm{T}_{2,\mathrm{rot}}=2/\Gamma_{P}=1.8$ ms. This relaxation time is not far from the measured Hahn-echo $T_{2}^{\mathrm{H}}=1.1$ ms for a P donor electron spin in enriched $^{28}$Si, believed to be limited primarily by the static magnetic field noise and thermal noise, and not due to the proximity to the oxide layers or other amorphous material \cite{Muhonen2014}. Therefore, it is reasonable to assume that the spin $T_{2}$ time in our device is Purcell limited, hence set by $T_{2,\mathrm{rot}}$. Note that for the Al/Bi set, the direction of the Stark shift must be such that the $\ket{F,m_{F}}=\ket{5,-5}\leftrightarrow\ket{4,-4}$ transition frequency, already reduced by a relatively small $B_{0}$ to lift the multiplet degeneracy, is further reduced to avoid exciting the higher frequency multiplet transitions. In this dispersive regime ($\Delta\gg g$), the microwave drive frequency is $\omega_{\mu\mathrm{w}}/2\pi=\omega_{\mathrm{s}}/2\pi+(2n_{\mathrm{lim}}+1)g^{2}/2\pi\Delta=5.614406$ GHz where photon number $n_{\mathrm{lim}}=\Delta^{2}/4g^{2}=400$ sets the maximum limit for the drive power \cite{PhysRevA.69.062320}, and is lower than the critical photon number $n_{\mathrm{crit}}\approx1800$ corresponding to the spiral inductor's critical current. The qubit rotation speed, at on-resonance photon number $n_{\mathrm{res}}\approx17\times10^{6}$ set by $n_{\mathrm{lim}}$ and $\omega_{\mu\mathrm{w}}$, is $f_{\mathrm{rot}}=g\kappa\sqrt{n_{\mathrm{res}}}/\pi\Delta=29$ MHz \cite{haroche1993fundamental}, corresponding to $N_{\pi}=2f_{\mathrm{rot}}T_{2,\mathrm{rot}}>10^{5}$ coherent $\pi$-rotations.

{\bf Readout:} The spin readout also occurs in the dispersive regime, where the cavity frequency depends on the spin state, giving rise to a $f_{sep}=g^2/\pi\Delta=12$ kHz separation frequency. This corresponds to a phase shift of $\phi=2\arctan(2g^{2}/\kappa\Delta)=10^{\circ}$, well above the measurement sensitivity usually considered to be $0.1^{\circ}$, and suggests an extremely high-fidelity readout when the resonator is driven at $\omega_0$. The measurement time to resolve $\phi$ is estimated to be $T_m=(2\kappa n_{\mathrm{read}}\theta_0^2)^{-1}= 3.1\ \mu$s where $\theta_0\equiv2g^{2}/\kappa\Delta$ \cite{PhysRevA.69.062320} and $n_{\mathrm{read}}=25$ is the readout photon number. In order to avoid the highly nonlinear response regime, the readout power must be limited such that $n_{\mathrm{read}}\lesssim\Delta^2/4g^2$. 

\section{V. Conclusion}

In summary, we have proposed and designed a novel device that enhances the coupling of a single-atom spin to the magnetic field of a circuit resonator by approximately 100 times compared to the previously proposed architectures that use coplanar resonators. This dramatic improvement is a result of using a low impedance, lumped element resonator design and a nanoscale spiral inductor geometry. We showed the possibility of entering the strong coupling regime necessary for practical purposes, {\it i.e.}, spectroscopic measurements and qubit realization. Using the well-established principles of cavity quantum electrodynamics, we showed that this large $g$ can lead to a significantly enhanced spin relaxation rate desired for qubit initialization, tens of megahertz spin rotation speed during manipulation without the need for an Oersted line, and superb dispersive readout sensitivity. Moreover, this architecture can be useful for coupling distant qubits using cavity photons for the realization of multi-qubit gates.  We emphasize that this large enhancement in the coupling rate is resulted directly from increasing the ac magnetic field at the spin, rather than by hybridizing the spin and charge states through, {\it e.g.}, a magnetic field gradient generated by local micromagnets.  In this way, we believe that a much larger spin-photon coupling rate can be achieved while avoiding a concomitant increase in the charge noise-induced decoherence rate \cite{Samkharadze1123,mi2018coherent}.

We also proposed a relatively simple proof-of-principle experiment which does not require single-donor implantation or e-beam lithography. This simplified scheme takes advantage of resonator's enhanced coupling to a spin {\it ensemble}, expected to result in the direct observation of the vacuum Rabi splitting to confirm strong coupling.

\section{acknowledgments}

The authors thank G. Bryant, D. Pappas, T. Purdy, M. Stewart, K. Osborn, J. Pomeroy, C. Lobb, A. Morello, C. Richardson, R. Murray and R. Stein for many useful discussions.

\section{Appendix: inductance simulations, magnetic field and coupling rate calculation}

The total inductance $L_{\mathrm{tot}}=L_{\mathrm{g}}+L_{\mathrm{p}}$ within the resonator consists of the geometric inductance $L_{\mathrm{g}}$ of the spiral inductor giving rise to the magnetic field $B_{\mathrm{ac}}$ that couples to the donor electron spin, and parasitic inductance $L_{\mathrm{p}}$ which does not contribute to $B_{\mathrm{ac}}$ and only limits it. $L_{\mathrm{g}}$ consists of the trace inductance $L_{\mathrm{t}}$ arising from the length of the spiral trace, and some mutual inductance $L_{\mathrm{M}}$ between the loops such that $L_{\mathrm{g}}=L_{\mathrm{t}}+L_{\mathrm{M}}$. The parasitic inductance $L_{\mathrm{p}}$ includes the kinetic inductance $L_{\mathrm{k}}$ of the spiral, the kinetic inductance $L_{\mathrm{C,k}}$ within the capacitor plates and the geometric self-inductance $L_{\mathrm{C,g}}$ of the capacitor such that $L_{\mathrm{p}}=L_{\mathrm{k}}+L_{\mathrm{C,k}}+L_{\mathrm{C,g}}$. Since $L_{\mathrm{p}}$ does not create magnetic fields at the location of the spin and only limits $I_{\mathrm{ac}}$ through $L_{\mathrm{tot}}$, we want to minimize it. 
In order to obtain an accurate estimation of $L_{\mathrm{g}}$ and $L_{\mathrm{p}}$, we performed calculations as well as software simulations. We study two device geometries, one using a single-layer spiral inductor and the other using a double-layer spiral (2S) inductor shown in Figs. \ref{layout}(c) and \ref{2slayout}, respectively. 

\begin{figure}
	\centering
	\includegraphics[width=2.5 in]{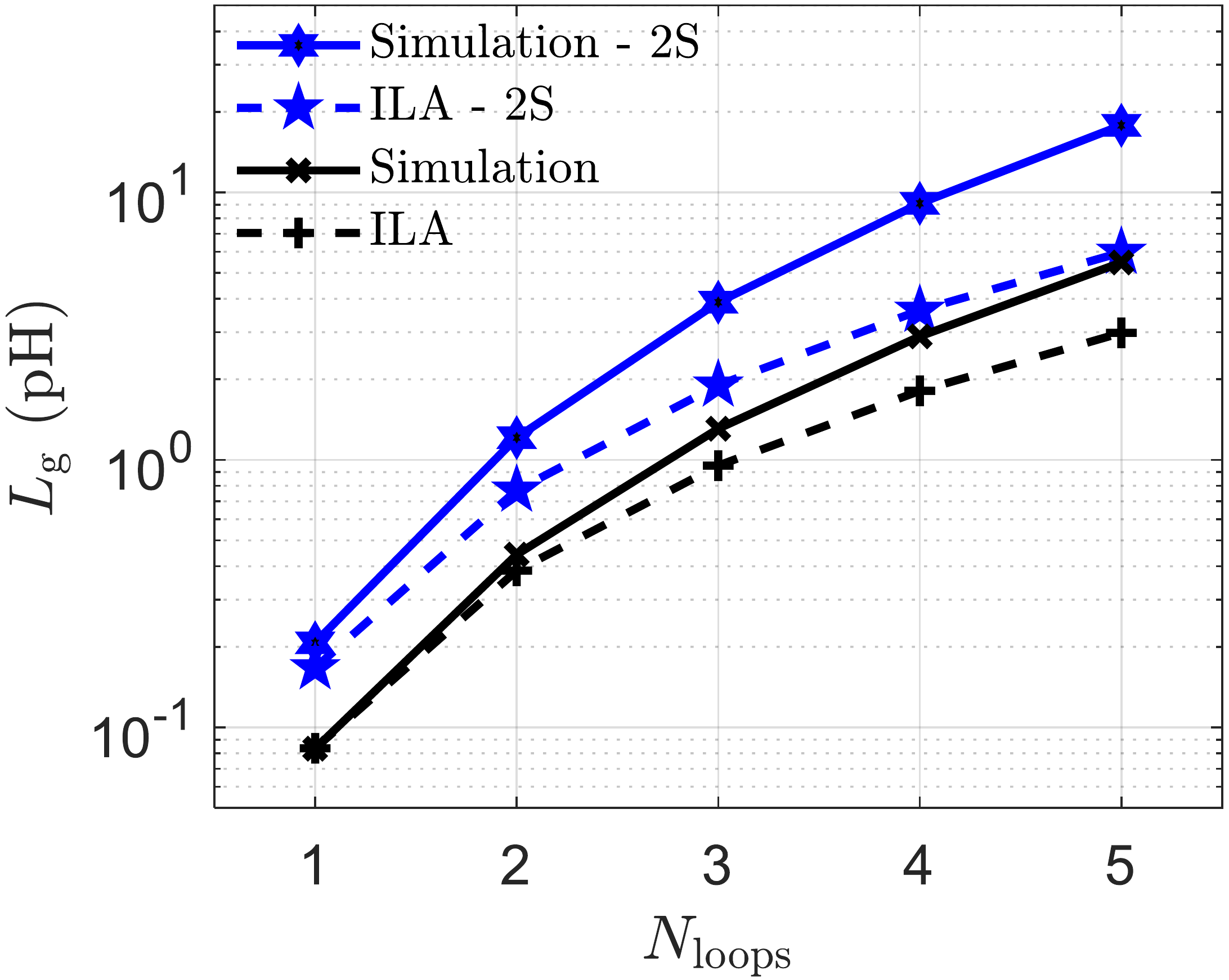}
	\caption{Comparison between the independent loop approximation (ILA) and FastHenry simulations of the spiral geometric inductance versus for different spiral loop counts $N_{\mathrm{loops}}$ for the single-spiral and double-spiral (2S) geometries.}
	\label{fig5}
\end{figure}

$L_{\mathrm{g}}$ is approximately calculated, and also separately simulated. In the calculations, for simplicity, we assume that each loop of the spiral inductor is perfectly circular and estimate $L_{\mathrm{g}}$ using the geometric mean distance (GMD) method to the second order in the ratio of the conductor diameter to the loop radius \cite{grover2004inductance}. This independent loop approximation (ILA) ignores the mutual inductance $L_{\mathrm{M}}$ within the spiral loops. However, one should note that $L_{\mathrm{M}}$ contributes to $B_{\mathrm{ac}}$ and is not parasitic. Nevertheless, in addition to the ILA, we also simulated the spiral geometry in FastHenry 3D inductance extraction program \cite{kamon1994fasthenry}, which accounts for the mutual inductances $L_{\mathrm{M}}$ within the spiral geometry. Figure \ref{fig5} shows a comparison between the simulation and the approximate calculation, where the latter ignores $L_{\mathrm{M}}$. Clearly, $L_{\mathrm{M}}$ constitutes a larger portion of $L_{\mathrm{g}}$ with increasing $N_{\mathrm{loops}}$, but is negligible up to $N_{\mathrm{loops}}\approx2,3$ where $g$ is optimum (see Fig. \ref{plots}(b)).
The total inductance $L_{\mathrm{tot}}$ in Fig. \ref{plots}(a) is plotted using $L_{\mathrm{g}}$ from FastHenry simulations. However, for simplicity, the ac field contributed by $L_{\mathrm{M}}$ was not taken into account in calculating $g$, resulting in an underestimated $g$ in Fig. \ref{plots}(b).

The kinetic inductance $L_{\mathrm{k}}$ of the spiral loop is caused by the kinetic energy of the quasi-particles within the superconductor and hence does not create any magnetic fields. In our design, $L_{\mathrm{k}}$ is the largest part of $L_{\mathrm{p}}$. Using the Cooper-pair density $n_{\mathrm{s}}$ and mass $m_{\mathrm{C}}$ for a superconducting material, the length of the superconducting line $l_{\mathrm{s}}$ and its cross-sectional area $A_{\mathrm{s}}$, one can estimate $L_{\mathrm{k}}=m_{\mathrm{C}}l_{\mathrm{s}}/4n_{\mathrm{s}}e^{2}A_{\mathrm{s}}$ \cite{tinkham1996introduction}, where $e$ denotes the electron charge. For the single-layer spiral Nb/P resonator with $1\leq N_\mathrm{loops}\leq5$ we obtain $0.6\ \mathrm{pH}\leq L_\mathrm{k}\leq 7\ \mathrm{pH}$, linearly proportional to the spiral trace length.  

The kinetic inductance within the capacitor plates, $L_{\mathrm{C,k}}\approx60$ fH, is negligible due to the large cross-sectional area of the plates. The geometric self-inductance of the capacitor, calculated using a stripline model, is $L_{\mathrm{C,g}}\approx62$ fH, suppressed by employing a relatively thin capacitor insulating layer (small $d_0$). To confirm this, we simulated the resonator geometry including the capacitor and found $L_{\mathrm{C,g}}=64$ fH, in agreement with the calculated value and also negligible for $N_{\mathrm{loops}}\gtrsim2$. 

In order to achieve the desired resonance frequencies with the relatively small inductances shown in Fig \ref{plots}(a), relatively large capacitances are required. By using a capacitor insulator thickness of $d_{0}=25$ nm in the single-atom device, we can keep the square-shaped capacitor dimensions below 200 $\mu$m, while significantly suppressing $L_{\mathrm{C,g}}$.

For the single-atom device, a simulation of $g$ as a function of the spiral trace width $w$, spacing $s$ and thickness $t$, was performed. By assuming $w=s=t$, the results showed that, for the Nb (Al) set, the optimum $g$ is obtained when $30\ \mathrm{nm}\lesssim w=s=t\lesssim35$ nm ($w=s=t\approx20$ nm), weakly depending on whether the single-layer spiral or the 2S geometry is used. If e-beam lithography resolution of 20 nm is implemented, the Al set can yield a spin-photon coupling rate of $g/2\pi=0.45$ MHz for $N_\mathrm{loops}=2$.


At the location of the spin in the single-atom device, $B_{\mathrm{ac}}$ is approximated as the sum of the magnetic field from all loops, {\it i.e}, $B_{\mathrm{ac}}=\sum\limits_{\mathrm{loops}}\mu_{0}I_{\mathrm{ac}}/2R'_{\mathrm{loop}}$ where $R'_{\mathrm{loop}}\equiv(d_{in}^2+4d_{s}^2)^{3/2}/2d_{in}^2$ is defined as a characteristic radius which accounts for the spin location with respect to the center of the loops and $d_{s}$ denotes the vertical spin displacement from this center (see Figs. \ref{layout}(c) and \ref{ensemble}). Note that the assumption of current flowing in the center of the spiral conductor underestimates $B_{\mathrm{ac}}$, $B_{\mathrm{ac},0}$ and $g$, because in reality the majority of current will flow closer to the superconductor-silicon interface due to the relatively large electric permittivity ($\epsilon_{\mathrm{r}}=12$) of silicon. 

For a better understanding of the dependence of $g$ on the number $N_{\mathrm{loops}}$ of the spiral loops, a naive picture may be helpful to the reader. To the first order for $N_{\mathrm{loops}}=1$, $I_{\mathrm{ac}}\propto1/\sqrt{L_{\mathrm{g}}+L_{\mathrm{p}}}$, and $L_{\mathrm{g}}\propto N_{\mathrm{loops}}^{2}\ln(N_{\mathrm{loops}})$, and $B_{\mathrm{ac},0},g\propto I_{\mathrm{ac}}\ln(N_{\mathrm{loops}})$, but $L_{\mathrm{p}}$ is a weaker function of $N_{\mathrm{loops}}$ than $L_{\mathrm{g}}$. This suggest that by increasing $N_{\mathrm{loops}}$, $B_{\mathrm{ac},0}$ and hence $g$ increase up to a point where $L_{\mathrm{g}}$ approaches $L_{\mathrm{p}}$, and $I_{\mathrm{ac}}$ begins to drop as $N^{\alpha}$, with $\alpha<-1$. Employing a lumped element design provides the required flexibility to reach this optimum $L_{\mathrm{g}}$ and the corresponding $N_{\mathrm{loops}}$.


DISCLAIMER: Certain commercial equipment, instruments, or materials (or suppliers, or software, ...) are identified in this paper to foster understanding. Such identification does not imply recommendation or endorsement by the National Institute of Standards and Technology, nor does it imply that the materials or equipment identified are necessarily the best available for the purpose.

\end{document}